\documentstyle[12pt]{article}
\begin{document}

\title{Quantum information processing and multi-qubit entanglement with Josephson-junction charge qubits in a thermal cavity}
\author{XuBo Zou, K. Pahlke and W. Mathis\\
Electromagnetic Theory Group at THT,\\
Department of Electrical Engineering, \\
University of Hannover, Germany}

\date{\today}
\maketitle
\begin{abstract}
We propose a scheme to implement the two-qubit quantum phase gate
with Josephson-junction charge qubits in a thermal cavity. In this
scheme, the photon-number-dependent parts in the time evolution
operator are canceled at the special time. Thus the scheme is
insensitive to the thermal field. We also demonstrate that the
scheme can be used to generate maximally entangled state of many
Josephson-junction charge qubits

\end{abstract}
Recently, much attention has been paid to the quantum computers,
which are based on the fundamental principle of quantum mechanics.
The existence of quantum algorithms for specific problems shows
that a quantum computer can in principle provide a tremendous
speed up compared to classical computers\cite{shor,gro}. This
discovery motivated an intensive research into this mathematical
concept which is based on quantum logic operations on multi-qubit
systems\cite{bare}. In order to implement this concept into a real
physical system, a quantum system is needed, which makes the
storage and the read out of quantum information and the
implementation of the required set of quantum gates possible. This
system should be scalable and the isolation of the system from the
environment should be very well in order to suppress decoherence
processes. Several physical systems were suggested to implement
the concept of quantum computing: cavity QED systems\cite{tc},
trapped ion systems\cite{cd}, nuclear magnetic resonance
systems\cite{cg}. These systems have the advantage of high quantum
coherence, but cannot be integrated easily to form large-scale
circuits. There exists better potential to realize large-scale
quantum computers by implementation of qubit in solid-state system
based on electron spins in quantum dots\cite{loss} or nuclear
spins of donor atoms in silicon\cite{ka}.

Recently, superconducting charge\cite{ma,na1,na2} and phase
qubits\cite{moo} have attracted much attention because of possible
large-scale integration and relatively high quantum coherence. In
this paper, we focus on the superconducting charge qubits. In the
experiment\cite{na1}, Nakamura et al have demonstrated the
coherent oscillations of Cooper pairs on a superconducting
Cooper-pair box. This corresponds to a rotation operation of a
single charge qubit. More recently, Pashkin et al observed the
quantum oscillations in two coupled charge qubits, which
demonstrated the feasibility of coupling of multiple charge qubits
\cite{na2}. In this experiment, the way of coupling Josephson
charge qubits is to connect two Cooper-pair box via a capacitor.
The disadvantage of this coupling is hard to directly couple two
distant charge qubits. In Ref\cite{ma}, a theoretical scheme was
proposed for coupling charge qubits in terms of the oscillator
modes in a LC circuit formed by an inductance and the qubit
capacitors. This scheme requires that the eigenfrequency
$\omega_{LC}$ of the LC circuit is much bigger than the quantum
manipulation frequencies, which not only makes quantum system
operate at a low speed but also limits the allowed number $N$ of
the qubits in the circuit because $\omega_{LC}$ scales with
$1/\sqrt{N}$. In Ref\cite{yang}, a fast scheme was proposed for
implementing quantum logic gate by placing three-level SQUID
qubits in a cavity.

In this paper we propose an alternative scheme to implement the
quantum-phase gate by placing two-level charge qubits in a cavity.
The distinct feature of the present scheme is that the
photon-number-dependent parts in the evolution operator are
canceled at the special time and two subsystem (cavity field and
charge qubits ) are disentangled. Due to this feature, the scheme
is insensitive to the thermal field. We also demonstrate that the
scheme can be used to generate maximally entangled states of many
charge qubits

We now consider the interaction of N charge qubits with a
single-mode cavity. The associated Hamiltonian of the system is
$$
H=H_{photon}+\sum_{j=1}^N H_{C}^j\eqno{(1)}
$$
where $H_{photon}=\omega a^{\dagger}a$ is the Hamiltonian of the
cavity mode, $a$ and $a^{\dagger}$ are the annihilation and
creation operators of the cavity field  of frequency $\omega$.
$H_{C}^j$ is the Hamiltonian of the $j$th charge qubit. The single
charge qubit is shown in Fig.1, which has been proposed in
Ref\cite{ma}. This charge qubit consists of two symmetric
Josephson junctions in a loop configuration, which can be tuned by
an external classical magnetic flux $\Phi_{c}$ which is controlled
by the current through the inductor loop. A controllable gate
voltage $V_g$ is coupled to the charge qubit via a gate capacitor
$C_g$. If the self-inductance of the loop is low, the charge qubit
is described by the Hamiltonian of the form
$$
H_{C}^j=4E_C(n_j-n_{g})^2-E_{J0}\cos\left(\Theta_j-\gamma_j\right
) -E_{J0}\cos\left (\Theta_j+\gamma_j\right ) \eqno{(2)}
$$
Here $E_C=e^2/2(C_g+2C_{J0})$ and $n_{g}=C_gV_g/2e$. $C_{J0}$ and
$E_{J0}$ are the capacitance and coupling energy of one Josephson
junction. In this paper, we assume that two Josephson junctions in
each charge qubit are same. $n_j$ is the number operator of
Cooper-pair charges on the island, $\Theta_j$ is the phase
difference across the Josephson junction and satisfies the
commutation relation $[n_j,\Theta_k]=-i\delta_{jk}$. The
interaction between the charge qubit and the cavity field is
contained in $
\gamma_j=\frac{\pi}{\Phi_0}\Phi_c+\frac{2\pi}{\Phi_0}\int_jA(x)\cdot{dl}
$, where $\Phi_{0}=hc/2e$ is the flux quantum and the $A(x)$ is
vector potential which arises from the electromagnetic field of
the normal mode of the cavity, and line integral is taken across
the junction. In the Coulomb gauge, this vector potential takes
the form
$A=\sqrt{hc^2/\omega{V}}(a+a^{\dagger})\hat{\epsilon}$\cite{samm},
where $\hat{\epsilon}$ is the unit polarization vector of the
cavity mode and $V$ is the volume of the cavity mode. We assumed
that the junction dimensions are much smaller than the wavelength
of the cavity mode, so that the cavity electric field is
approximately uniform within the junction. We define the coupling
constant
$g=2\sqrt{2}e\sqrt{\frac{\pi}{\hbar\omega{V}}}\hat{\epsilon}\cdot\vec{l}$
such that
$\frac{2\pi}{\Phi_0}\int_jA(x)\dot{dl}=g(a+a^{\dagger})$. The
$\vec{l}$ is the thickness of the insulating layer in one
junction.

Now we consider the system in which the charging energy is much
larger than the Josephson coupling energy $E_C\gg E_{J0}$.  In
this regime, a convenient basis is formed by the charge states,
parametrized by the number of Cooper pairs on the island. In this
basis, the Hamiltonian(1) reads
$$
H=\omega a^{\dagger}a+\sum_{j=1}^N\sum_{n_j=0}^{\infty}[
4E_C(n_j-n_{g})^2|n_j\rangle\langle n_j|
$$
$$
-E_{J0}\cos\left [ \frac{\pi}{\Phi_0}\Phi_c+g(a+a^{\dagger})\right
](|n_j\rangle\langle n_j+1|+|n_j+1\rangle\langle n_j|)] \eqno{(3)}
$$
We concentrate on the dimensionless gate charge $n_g$ near the
degeneracy point 1/2 ($n_g\approx1/2$), where only two charge
states, say $n_j=0$ and $n_j=1$, play a role, all the other charge
states, having a much higher energy, can be ignored. In this case,
the Hamiltonian(3) can be written as follows
$$
H=\omega{a^\dagger
a}+\Delta\sum_{j=1}^N\sigma_{jz}-E_{J0}\cos\left [
\frac{\pi}{\Phi_0}\Phi_c+g(a+a^{\dagger})\right
]\sum_{j=1}^N\sigma_{jx}\eqno{(4)}
$$
where $\Delta=2E_C(2n_g-1)$, $\sigma_{jz}$ and $\sigma_{jx}$ are
Pauli operators of the $j$th charge qubit and the charge states
$n_j=0$ and $n_j=1$ correspond to the spin basis states. In the
following, We assume that the the dimensionless gate charge $n_g$
is tuned to be $1/2$ and the coupling constant between cavity and
junction is weak $g\sqrt{\bar{n}+1}\ll1$, here $\bar{n}$ is mean
photon number of cavity mode. Under these conditions, we perform
the expansion
$$
\cos\left [ \frac{\pi}{\Phi_0}\Phi_c+g(a+a^{\dagger})\right
]\approx\cos\left [ \frac{\pi}{\Phi_0}\Phi_c\right
]-g(a+a^{\dagger})\sin\left [ \frac{\pi}{\Phi_0}\Phi_c\right ]
\eqno{(5)}
$$
and the Hamiltonian(4) can be written as
$$
H=\omega{a^{\dagger}a}-E_{J0}[\cos\theta-g(a+a^{\dagger})\sin\theta]J_x
\eqno{(6)}
$$
where $\theta=\frac{\pi}{\Phi_0}\Phi_c$ and collective spin
operator $J_{x}=\sum_{i=1}^{N}\sigma_{ix}$. The parameter $\theta$
is controllable. The exact time evolution operator of
Hamiltonian(6) is
$$
U(t)=\exp[-i\omega{t}{a^{\dagger}a}-iE_{J0}gt\sin\theta(a+a^{\dagger})J_x]\exp(iE_{J0}t\cos\theta
J_x) \eqno{(7)}
$$
which can be rewritten in the form
$$
U(t)=\exp\left [iJ_x^2\left
(\frac{g^2E_{J0}^2\sin^2\theta{t}}{\omega}-\frac{g^2E_{J0}^2\sin^2\theta\sin\omega{t}}{\omega^2}\right
)\right ] \exp(-i\omega{a^{\dagger}a}t)$$ $$ \times\exp\left
[\frac{gE_{J0}\sin\theta}{\omega}J_x\left
(a^{\dagger}(e^{i\omega{t}}-1)-a(e^{-i\omega{t}}-1)\right )\right
]\exp(iJ_xE_{J0}\cos\theta{t}) \eqno{(8)}
$$
If we choose the interaction time $\tau$ to satisfy the condition
$\tau=2\pi/\omega$, the time evolution operator reduces to
$$
U(\tau)=\exp\left [\frac{2i\pi
g^2E_{J0}^2\sin^2\theta}{\omega^2}J_x^2\right ]\exp\left
[\frac{2i\pi E_{J0}\cos\theta}{\omega}J_x\right ]\eqno{(9)}$$ It
is easy to see, at the time $\tau$, two subsystem are
disentangled. The cavity mode is returned to its original state,
be it the ground state or any excited state, and we are left with
an internal state evolution, which is independent of the cavity
state. For two charge qubits, we choose the parameters $g$,
$E_{J0}$, $\theta$ and $\omega$ to satisfy
$$
\cos\theta=\frac{g}{\sqrt{g^2+4}} \quad \mbox{and}\quad
\frac{\omega}{E_{J0}}=\frac{8g}{\sqrt{g^2+4}}\eqno{(10)}$$ In this
case, the time evolution operator becomes
$$
U(\tau)=\exp\left [\frac{i\pi}{8}(J_x^2+2J_x)\right ]\eqno{(11)}$$
 It is easy
to check that this time evolution operator represents a quantum
phase gate
$$
|-\rangle_1|\pm\rangle_2\longrightarrow|-\rangle_1|\pm\rangle_2$$
$$
|+\rangle_1|-\rangle_2\longrightarrow|+\rangle_1|-\rangle_2$$
$$
|+\rangle_1|+\rangle_2\longrightarrow-|+\rangle_1|+\rangle_2
\eqno{(12)}
$$
in the basis states
$|\pm\rangle_i=\frac{1}{\sqrt{2}}(|0\rangle_i\pm|1\rangle_i)$.

We now turn to the problem of generating an entangled state of N
charge qubits
$$
|\Psi>=\frac{1}{\sqrt{2}}[e^{i\varphi_g}|00\cdots{0}>+e^{i\varphi_e}|11\cdots{1}>]\,.
\eqno{(13)} $$ irrespective of N even or odd. Quantum states of
this kind were used to improve the frequency standard. Several
schemes \cite{bo,ms} were proposed to generate this kind of
quantum states. We assume that the system was initially prepared
in the ground state $|000\cdots>$. If $N$ is even, we choose the
parameters to satisfy $\theta=\pi/2$ and $gE_{J0}/\omega=1/4$. The
time evolution operator will at the time $\tau=2\pi/\omega$ is
$U(\tau)=\exp\left [\frac{i\pi}{8}J_x^2\right ]$. In Ref.\cite{ms}
such a kind of time evolution operator was used to generate
quantum states of the form(16) with $\varphi_g=-\frac{\pi}{4}$ and
$\varphi_e=\frac{\pi}{4}+\frac{N\pi}{2}$. In the case of odd
numbers $N$ of charge qubits our scheme makes the generation of
maximally entangled quantum states(13) possible by using the time
evolution operator (11).

In summary, we proposed a scheme to implement a quantum phase gate
or to entangle quantum states of $N$ charge qubits by placing
qubits in a cavity. We calculate the exact time evolution and
demonstrate how the two subsystems are disentangled at particular
time $\tau=2\pi/\omega$. Thus the photon-number dependent parts in
the evolution operator are canceled and scheme is insensitive to
the thermal field. Thus, the quantum gate operation speed can be
greatly increased, which is important in view of decoherence.

\begin{flushleft}

{\Large \bf Figure Captions}

\vspace{\baselineskip} {\bf Figure 1.} Single charge qubit
consists of two symmetric Josephson junctions in a loop
configuration, which can be tuned by an external classical
magnetic flux $\Phi_{c}$

\end{flushleft}
\end{document}